\documentstyle[prd,aps, 
preprint,   
]{revtex}  

\begin{document}  
  
\draft  
\preprint{\begin{tabular}{r}  
{\bf (hep-ph/0012046)}\\  
KIAS-P00063\\  
YUMS 00-06\\    
\end{tabular}}  
  
\title{Majorana Neutrino Masses and Neutrino Oscillations}  
  
\author{  
Sin Kyu Kang $^{a,}$\footnote{skkang@kias.re.kr}~~~ and ~~~  
C. S. Kim $^{b,c,}$\footnote{kim@kimcs.yonsei.ac.kr,~~  
http://phya.yonsei.ac.kr/\~{}cskim/}  
}  
\address{  
$^a$ School of Physics, Korea Institute for Advanced Study,   
Seoul 130-012, Korea \\  
$^b$ Physics Department and IPAP, Yonsei University, Seoul 120-749, Korea \\   
$^c$ Department of Physics, University of Wisconsin, Madison, WI 53706, USA   
 }  
\maketitle  
  
\begin{abstract}  
\noindent  
We examine some patterns of Majorana neutrino mass matrix which is compatible  
with the phenomenological lepton flavor mixing matrix and non-observation  
of neutrinoless double beta decay.   
Imposing $(M_{\nu})_{ee}=0$  
for the Majorana neutrino mass matrix in the leading order, we obtain  
a relationship between the solar mixing angle and   
the neutrino masses $m_1$ and $m_2$. Additional possible texture  
zeros are assigned to the mass matrix so as for the nonvanishing  
$\theta_{13}$ to be predictable in terms of neutrino masses.  
We also show how three neutrino masses can be predicted  
from the solar mixing angle and the experimental results of 
$\Delta m^2_{sol}$ and $\Delta m^2_{atm}$   in this framework, 
and we discuss naturality of the
forms of the mass matrix  found in our work. 
\end{abstract}  
  
  
  
  
Although the lepton flavor mixing matrix can be constructed  
based on the neutrino oscillation experimental results, the origin  
of the lepton flavor mixing, the neutrino masses, and their  
hierarchical patterns are yet to be understood.  
As an attempt toward the understanding of the neutrino masses and   
flavor mixing patterns, the mass matrix ansatz has been studied by many   
authors  \cite{texture1}.  
In this approach, the flavor mixing elements are not independent parameters  
but are presented in terms of mass eigenvalues. The special patterns   
of the flavor mixing and masses can be achieved by imposing some flavor  
symmetry or taking texture zeros in the mass matrix as much as possible.  
  
As is well known,  current data from the atmospheric \cite{superkam} and  
solar  neutrino experiments \cite{solar} provide convincing evidence  
that neutrinos may have nonzero masses and oscillate, and  terrestrial  
neutrino experiments \cite{lsnd,chooz,verde} also lead to meaningful  
constraints on neutrino masses and mixing:  
\begin{itemize}  
\item   
(a) The atmospheric neutrino experimental data  indicates the near maximal  
mixing between $\nu_{\mu} $ and $\nu_{\tau}$, $\sin^2 2\theta_{atm}\geq 0.8$,  
with a mass squared difference  
$\Delta m^2_{atm} \simeq (0.5 \sim 6)\times 10^{-3}~\mbox{eV}^2$ at $90\%$CL  
\cite{superkam}. The best fit occurs at $(\sin^2 2\theta_{atm}=1.0,  
\Delta m^2_{atm}=2.2\times 10^{-3}~\mbox{eV}^2)$.  
\item  
(b) The solar neutrino anomaly can be explained through matter enhanced  
neutrino oscillation \cite{msw} if  
$ \Delta m^2_{sol} \simeq (0.4 \sim 1)\times 10^{-5} ~\mbox{eV}^2$ and  
$\sin^2 2 \theta _{sol}\simeq (0.1 \sim 1)\times 10^{-2}$  
(small angle MSW (SMA)), or  
$\Delta m^2_{sol} \simeq (1.5 \sim 10)\times 10^{-5} ~\mbox{eV}^2$,  
$\sin^2 2 \theta _{sol}\geq 0.6$ (large angle MSW (LMA)),   
$\Delta m^2_{sol}\sim  
10^{-7} ~\mbox{eV}^2, \sin^2 2 \theta_{sol}\sim 1.0$ (LOW solution)  
\cite{LOW} and through long-distance  vacuum oscillation (VO) if  
$\Delta m^2_{sol} \simeq 10^{-10} ~\mbox{eV}^2$,  
$\sin^2 2 \theta _{sol}\geq 0.7$.  
\item  
(c) Moreover, the CHOOZ experimental results can constrain $\nu_e - \nu_{x}$  
oscillation with  $\Delta m^2_{atm}\geq 10^{-3}  
~\mbox{eV}^2$ \cite{chooz} but gives no limit for $\Delta m^2_{atm}< 10^{-3}
~\mbox{eV}^2$, and the recent Palo Verde reactor experiment
also indicates no atmospheric $\nu_e-\nu_{x}$ oscillation  
for $\Delta m^2 \geq 1.12\times 10^{-3}$ and  
for $\sin^2 2\theta \geq 0.21$(for large $\Delta m^2$)~\cite{verde}.  
{}From those reactor experiments, we can obtain a constraint on the  
magnitude of $U_{e3}$, which is turned out to be small, {\it i.e.},  
$|U_{e3}|\leq 0.22$.   
\item  
(d) In the case of the LSND experiment at LANL \cite{lsnd}, 
a evidence for $\overline{\nu_{\mu}} \rightarrow \overline{\nu_e}$ oscillation  
has been reported. However, since the LSND result has not  
yet been independently confirmed by other similar experiments, we do not 
include it in our analysis.  \item  
(e) Now we assume that there are only three active neutrinos with  
Majorana masses.  For convenience, let us adopt the following convention.  
The heaviest neutrino mass eigenstate responsible for the atmospheric  
neutrino anomaly is taken to be $\nu_3$, whereas those responsible  
for the solar neutrino problem are $\nu_1$ and $\nu_2$.  
Then, the mass squared  
differences between two atmospheric neutrinos and two solar neutrinos become  
$\Delta m^2_{atm}\simeq \Delta m^2_{32} \simeq \Delta m^2_{31}$ and  
$\Delta m^2_{sol}\simeq \Delta m^2_{21}$, respectively.  
\end{itemize}  
Using the above experimental constraints on the neutrino masses and mixing   
angles, one can construct the phenomenological lepton flavor mixing matrix,  
as follows:  
\begin{itemize}  
\item  
(i) Since the best fitted value of the Super-Kamiokande data for the  
atmospheric neutrino mixing angle corresponds to the maximal mixing, we take  
$\theta_{23}=\pi/4$. However, since there are two possibilities for the solar   
neutrino mixing angle $\theta_{12}$ as shown in the above, we present  
corresponding elements of the mixing matrix in terms of   
$\theta_{12}\equiv \theta$.   
\item  
(ii) As for the mixing angle $\theta_{13}$ which is related to  
$U_{e3}$, there is only upper bound on its value as presented above.   
As shown in Ref. \cite{fit}, the fitted values for the oscillation  
amplitude for solar neutrinos are not greatly affected by the particular  
value of $s_{13}$ in this case, thus we take $\sin \theta_{13}\equiv  
\epsilon$ and $\cos \theta_{13} \sim 1$ in the leading order.   
\end{itemize}  
Then in general the  
lepton  mixing matrix in the standard parametrization has the form in the  
leading order \cite{akh},   
\begin{eqnarray}   
U &=&  \pmatrix{c_{13}c_{12} & c_{13}s_{12} & s_{13} \cr  
-c_{23}s_{12}-s_{13}s_{23}c_{12} & c_{23}c_{12}-s_{13}s_{23}s_{12} &   
          c_{13}s_{23}\cr  
s_{23}s_{12}-s_{13}c_{23}c_{12} & -s_{23}c_{12}-s_{13}c_{23}s_{12} &  
c_{13}c_{23}} \nonumber \\  
 &=& \pmatrix{ c & s & \epsilon \cr  
 -\frac{1}{\sqrt{2}}(s+c\epsilon) & \frac{1}{\sqrt{2}}(c-s\epsilon) &   
  \frac{1}{\sqrt{2}} \cr  
  \frac{1}{\sqrt{2}}(s-c\epsilon) & -\frac{1}{\sqrt{2}}(c+s\epsilon) &   
  \frac{1}{\sqrt{2}} } , \label{lepm}  
\end{eqnarray} \\  
where $c=\cos\theta_{12}, s=\sin\theta_{12}$, and the neutrino flavor  
basis is $(\nu_e, \nu_{\mu}, \nu_{\tau})$.  
Here, we assume that there is no CP violation in the lepton sector.   
  
In this Letter, we will examine some   
patterns of Majorana  
neutrino mass matrix which is compatible with the above lepton mixing matrix  
and reflects the predictable framework of neutrino masses.  
Recently, there has been much work suggesting various  
textures of neutrino masses by using  some phenomenological ansatz \cite{ansatz}  
or some symmetry arguments \cite{symm}, such as $SO(10), SO(3), U(2), U(1)$, 
etc.   Here we will  take the approach to require texture 
zeros from the appropriate experimental  observations, instead of imposing 
specific flavor symmetry in the neutrino sector.    The general form of mass 
matrix presented in terms of three neutrino mass   eigenvalues will be 
provided with the help of the lepton mixing matrix.  Motivated by 
non-observation of neutrinoless double beta decay,  we impose 
$(M_{\nu})_{ee}=0$ for the Majorana neutrino mass matrix {\it in the  leading 
order}, which in turn makes the solar mixing angle simply related to  the 
ratio $m_1/m_2$.  This is consistent  as long as  
$\epsilon^2<<m_2\sin^2\theta_{12}/m_3$.  Additional possible  texture zeros 
will be assigned to the mass matrix    so as for the mixing parameter 
$\epsilon$ to be predictable in terms   of neutrino mass eigenvalues.   
We will also show that three neutrino mass eigenvalues  
can be calculated from the relation for the solar mixing angle and  
the experimental results of $\Delta m_{sol}^2$ and $\Delta m_{atm}^2$.  
  
A strong constraint on some element of Majorana neutrino mass matrix can  
come from the experimental results of neutrinoless double beta decay,  
whose non-observation might serve as a texture zero for the leading   
order mass matrix.  
If massive neutrinos are Majorana particles,   
the matrix element of the neutrinoless double  
beta decay is proportional to the effective Majorana mass  
\begin{equation}  
|<m_{\nu}>|=|\sum_{i}m_iU_{ei}^2|,  
\end{equation}  
where $U$ is the lepton mixing matrix that connects the flavor neutrino   
eigenstates $\nu_{\alpha L} ~(\alpha=e,\mu, \tau)$ to the mass  
eigenstates $\nu_i ~(i=1,2,3)$ through the relation  
\begin{equation}  
\nu_{\alpha L} = \sum_{i}U_{\alpha i}\nu_{iL}.  
\end{equation}  
This effective Majorana mass is equal to the absolute value of  
the element $(M_{\nu})_{ee}$ of the mass matrix in the charged lepton  
flavor basis, {\it i.e.} the mass matrix for the charged leptons  in its  
diagonal basis.  
The current experimental upper bound on $|<m_{\nu}>|$ is given by  
\cite{bb0n},  
\begin{equation}  
|<m_{\nu}>|\leq 0.2 ~~\mbox{eV} ~~(90\% \mbox{C.L.}).  
\end{equation}  
The GENIUS experiment is expected to be sensitive to   
$|<m_{\nu}>|$ as low as 0.01 eV  or even 0.001 eV \cite{genius,bb02}.  
Thus, the magnitude of the element $(M_{\nu})_{ee}$ might be strongly  
constrained by the experimental results of neutrinoless double beta decay.  
Although it is not yet proved, it is possible to enforce $(M_{\nu})_{ee}  
=0$ for some special pattern of neutrino mixing \cite{bimax1}.  
In this paper, we require $(M_{\nu})_{ee}=0$ in the leading order,   
from which the solar mixing angle is simply related to the neutrino mass   
ratio $m_1/m_2$. We note that although $(M_{\nu})_{ee}=0$   
in the leading order,   
there is nonvanishing very small next leading contribution to   
$(M_{\nu})_{ee}$   
due to nonzero $\epsilon$ parameter whose magnitude is proportional to   
$\epsilon^2$.  
  
The Majorana neutrino mass matrix in the charged lepton flavor basis can be  
given by $M_{\mu}=U\cdot D\cdot U^{T}$. The diagonal matrix can be written as  
$diag[m_1e^{i\alpha}, m_2e^{i\beta}, m_3]$, where $m_i(i=1,2,3)$ is  
positive definite.  
For CP conserving case, the phases $\alpha$ and $\beta$ are taken to be  
either $\pi$ or $0$. Thus, we can consider the possible three cases:  
\begin{itemize}  
\item  
{\bf Case (1)} $M_{\nu}=U\cdot diag[-m_1, m_2, m_3]\cdot U^{T}$,  
\item  
{\bf Case (2)} $M_{\nu}=U\cdot diag[m_1, -m_2, m_3]\cdot U^{T}$,   
\item  
{\bf Case (3)} $M_{\nu}=U\cdot diag[\pm m_1, \pm m_2, m_3]\cdot U^{T}$.  
\end{itemize}  
{\bf Case (1):} Keeping the $\epsilon^2$ order, the neutrino mass matrix  
is presented by  
\begin{eqnarray}\label{massm1}  
M_{\nu} &=& \pmatrix{  
 w_1 & \frac{1}{\sqrt{2}}[\epsilon(m_3-w_1)+m_{+}cs]  
 & \frac{1}{\sqrt{2}}[\epsilon(m_3-w_1)-m_{+}cs]  \cr  
  \frac{1}{\sqrt{2}}[\epsilon(m_3-w_1)+m_{+}cs] &  
 \frac{1}{2}(m_3+w_2-2m_{+}cs\epsilon) &  
 \frac{1}{2}(m_3-w_2) \cr  
 \frac{1}{\sqrt{2}}[\epsilon(m_3-w_1)-m_{+}cs]  &  
 \frac{1}{2}(m_3-w_2) &  
 \frac{1}{2}(m_3+w_2+2m_{+}cs\epsilon) } \nonumber \\  
& & + \epsilon^2 \pmatrix{  
m_3 - w_1 & -\frac{cs}{2\sqrt{2}}m_+ & \frac{cs}{2\sqrt{2}}m_+ \cr  
-\frac{cs}{2\sqrt{2}}m_+ & -\frac{1}{2}(m_3-w_1) & -\frac{1}{2}(m_3-w_1)   \cr  
\frac{cs}{2\sqrt{2}}m_+ & -\frac{1}{2}(m_3-w_1) & -\frac{1}{2}(m_3-w_1) } ,  
\end{eqnarray}   
where $w_1=-m_1c^2+m_2s^2, w_2=-m_1s^2+m_2c^2$ and $m_{+}=m_1+m_2$.   
{}From the condition $(M_{\nu})_{ee}=0$ in the leading order,  
the following relation comes out  
\begin{eqnarray}  
\tan\theta_{12} &=& \tan\theta_{sol} =\sqrt{\frac{m_1}{m_2}} .  
\end{eqnarray}    
Thus, in this ansatz,  
the solar neutrino mixing pattern is attributed to the ratio $m_1/m_2$.  
For the hierarchy $m_1,m_2<<m_3$, the natural choice for two additional   
texture zeros of the mass matrix  
in the leading order would be [$(M_{\nu})_{e\tau}, (M_{\nu})_{\tau e}$]   
elements,  which lead to the following relation;  
\begin{eqnarray}  
\epsilon &=& \frac{\sqrt{m_1m_2}}{m_3} .  
\end{eqnarray}    
{}From the CHOOZ experimental results,   
$U_{e3}$ can be constrained and we obtain    
\begin{equation}\label{choo}  
|U_{e3}|=\epsilon =\frac{\sqrt{m_1m_2}}{m_3} < 0.22 .  
\end{equation}   
Then, we are led to the leading order neutrino mass matrix in the charged   
lepton basis presented in terms of the three neutrino mass eigenvalues   
\begin{eqnarray}  
M_{\nu}= m_3\pmatrix{  
 0 & \sqrt{2}\epsilon & 0 \cr  
 \sqrt{2}\epsilon &   
  \frac{1}{2}\left(1+\frac{m_2-m_1}{m_3}\right) &  
  \frac{1}{2}\left(1-\frac{m_2-m_1}{m_3}\right) \cr  
0 &  \frac{1}{2}\left(1-\frac{m_2-m_1}{m_3}\right) &   
  \frac{1}{2}\left(1+\frac{m_2-m_1}{m_3}\right)   
 } + {\cal O}(\epsilon^2).  
\end{eqnarray}   
We note that this form of mass  matrix is similar pattern of the neutrino  
mass matrix presented in \cite{kkk}.  
{}From Eq. (\ref{choo}), we see that only the hierarchy $m_1,m_2 << m_3$   
is relevant for this form of mass matrix to be consistent  
with the experimental results.  
In this ansatz, the maximal mixing of solar neutrino oscillation is attributed  
to almost degenerate $\nu_1$ and $\nu_2$, while the small mixing is  
achieved by the hierarchy of $m_1$ and $m_2$.  
For the inverted hierarchy, $m_1\sim m_2 >> m_3$, one can have another form  
of mass matrix with three texture zeros by taking $m_3\simeq m_2-m_1$:  
\begin{eqnarray}  
M_{\nu} &\simeq& \pmatrix{  
 0 & \frac{1}{\sqrt{2}}\left(\epsilon m_3 + \sqrt{m_1m_2}\right) &  
  \frac{1}{\sqrt{2}}\left(\epsilon m_3 - \sqrt{m_1m_2}\right) \cr  
  \frac{1}{\sqrt{2}}\left(\epsilon m_3 + \sqrt{m_1m_2}\right) &  
m_3-\epsilon\sqrt{m_1m_2} & 0 \cr  
  \frac{1}{\sqrt{2}}\left(\epsilon m_3 - \sqrt{m_1m_2}\right) &  
0 & m_3-\epsilon\sqrt{m_1m_2}   } \\  
& &+ {\cal O}(\epsilon^2). \nonumber  
\end{eqnarray}  
However, in this case, the value of $\epsilon$ is not predicted.  
If the magnitude of $\epsilon$ is  
taken to be negligibly small, the form of mass matrix indicates nearly  
pseudo-Dirac neutrinos.  
  
\noindent  
{\bf Case (2):} Keeping only to ${\cal O}(\epsilon)$, we have  
\begin{eqnarray}\label{massm2}  
M_{\nu} &=& \pmatrix{  
 w_1 & \frac{1}{\sqrt{2}}[\epsilon(m_3-w_1)-m_{+}cs]  
 & \frac{1}{\sqrt{2}}[\epsilon(m_3-w_1)+m_{+}cs]  \cr  
  \frac{1}{\sqrt{2}}[\epsilon(m_3-w_1)-m_{+}cs] &  
 \frac{1}{2}(m_3+w_2+2m_{+}cs\epsilon) &  
 \frac{1}{2}(m_3-w_2) \cr  
 \frac{1}{\sqrt{2}}[\epsilon(m_3-w_1)+m_{+}cs]  &  
 \frac{1}{2}(m_3-w_2) &  
 \frac{1}{2}(m_3+w_2-2m_{+}cs\epsilon) }  \\   
& & +{\cal O}(\epsilon^2)~, \nonumber  
\end{eqnarray}   
where $w_1=m_1c^2-m_2s^2, w_2=m_1s^2-m_2c^2$ and $m_{+}=m_1+m_2$.   
The next leading order contribution $O(\epsilon^2)$ is similar to  
the {\bf Case (1)}.  
We are also led to the relation Eq. (6) from the condition of no neutrinoless   
double beta decay in the leading order.  
The natural choice for two additional texture zeros in this case is to take  
($M_{e\mu}, M_{\mu e}$) elements  to be zero  
which lead to the same relation as Eq. (7).  
Then, this form of mass matrix become triangular type of mass matrix:  
\begin{eqnarray}  
M_{\nu}= m_3\pmatrix{  
 0 & 0 & \sqrt{2}\epsilon  \cr  
 0 &  
  \frac{1}{2}\left(1-\frac{m_2-m_1}{m_3}\right) &  
  \frac{1}{2}\left(1+\frac{m_2-m_1}{m_3}\right) \cr  
 \sqrt{2}\epsilon &   
  \frac{1}{2}\left(1+\frac{m_2-m_1}{m_3}\right) &   
  \frac{1}{2}\left(1-\frac{m_2-m_1}{m_3}\right)   
 } + O(\epsilon^2).  
\end{eqnarray}   
  
\noindent  
As one can see, the {\bf Case (3)} leads to the nonzero effective Majorana  
mass in the leading order. This is incompatible with our ansatz of  
no neutrinoless double beta decay in the leading order.  
Thus, we do not consider this case any more.  
  
Now, let us demonstrate how neutrino masses can be determined from the   
above results.  The numerical values of the mass squared differences   
$\Delta m^2_{21}$ and $\Delta m^2_{32}$ can be obtained from  the   
experimental results of   
$\Delta m^2_{sol}$ and $\Delta m^2_{atm}$, respectively.  
Since the mixing angle $\sin^2 2\theta_{sol}$  is related to the mass   
eigenvalues $m_1$ and $m_2$ through the relation (6), combining this with  
the numerical value of $\Delta m_{21}^2$, one can first  
determine the numerical values of $m_1$ and $m_2$. Then, the mass  
eigenvalue $m_3$ is determined from $\Delta m^2_{atm}=\Delta m^2_{32}$.  
In this way, one can get  
possible ranges of three neutrino masses.  
However, since there are two possibilities for the  mixing angle   
$\theta_{12}$   
corresponding to two types of the solar neutrino mixing,   
we have to consider two cases. \\  
  
\noindent  
{\it \bf (A) Small mixing angle solution}:   
{}From $\sin^2 2\theta_{sol}\simeq 10^{-2}$,  
we obtain the mass ratio $m_1/m_2\simeq 0.01$ which implies  
that the mass hierarchy $m_1<<m_2$ is required. Combining this with  
the experimental results for $\Delta m^2_{sol}$ and $\Delta m^2_{atm}$,  
we obtain   
\begin{eqnarray}  
(m_1, m_2, m_3)=[ (0.5 \sim 8)\times 10^{-6},~~   
(2 \sim 3)\times 10^{-3},~~  
0.05]~~~\mbox{eV}.  
\end{eqnarray}   
{}From these results, we can obtain $\epsilon=(0.7\sim 3)\times 10^{-3}$, which  
is consistent with the CHOOZ and Palo Verde experimental results. \\  
{\it \bf (B) Large mixing angle solutions}:   
For large mixing of the solar neutrinos, Eq. (6) leads to   
$m_1\simeq m_2$.  We note that the exact maximal mixing for the solar   
neutrino oscillation is not realistic for this approach.  
Taking some value in the allowed region of  
$\sin^22\theta_{sol}$ except $1$, one can determine three neutrino  
mass eigenvalues in the same way as the SMA case.  
For example, if   
$$(\sin^2 2\theta_{sol},~~~ \Delta m^2_{sol},~~~ \delta m^2_{atm})$$   
are taken to be   
$$(0.9,~~~ 10^{-4}~(10^{-10})~\mbox{eV}^2,~~~ 0.0022~\mbox{eV}^2),$$  
respectively for LMA (VO),  
the allowed neutrino masses are then give by  
\begin{eqnarray}  
(m_1, m_2, m_3) = \pmatrix{6\times 10^{-3} & 1\times 10^{-2} & 0.05 &   
(\mbox{LAM})\cr 6\times 10^{-6} & 1\times 10^{-5} & 0.05 & (\mbox{VO})}  
~\mbox{eV}.  
\end{eqnarray}   
The prediction of $\epsilon$ is 0.17 and $2\times 10^{-4}$ for  
LMA and VO, respectively.  
Those results are also consistent with the experimental bounds.  
  
Based on the above numerical results, we can estimate the possible  
effective Majorana mass arising in the next leading order.  
{}From Eq. (5), it is given by $\epsilon^2 m_3$ and numerically  
$(0.2 \sim 5)\times 10^{-7}$ for SMA and  
$1.5 \times10^{-3} (2\times 10^{-9})$ for LMA  
(VO) with $\sin^22\theta_{sol}=0.9$.  
Those values are far below the current experimental bound given by Eq. (4).  
In particular, a new Heidelberg experimental proposal (GENIUS) will allow to   
increase the sensitivity for Majorana neutrino masses   
from the present level of  
0.1 eV down to 0.01 or even 0.001 eV. In an extended experiment using  
10 tons of $^{76}$Ge, a sensitivity of 0.001 eV could be reached \cite{bb02}.  
Thus, we expect that the test of the prediction of the possible  
 effective Majorana mass for LMA would be accessible in near future.  
  
Now, let us take into account the limit of $\epsilon=0$.  
Requiring no neutrinoless double beta decay, the mass matrix  
takes the form for the mass eigenvalues $(-m_1, m_2, m_3)$:  
\begin{eqnarray}  
M_{\nu}=\pmatrix{0 & -\sqrt{\frac{m_1m_2}{2}} & \sqrt{\frac{m_1m_2}{2}} \cr  
-\sqrt{\frac{m_1m_2}{2}} & \frac{1}{2}(m_3 + m_1 - m_2)  
& \frac{1}{2}(m_3 - m_1 + m_2) \cr  
\sqrt{\frac{m_1m_2}{2}} & \frac{1}{2}(m_3 - m_1 + m_2)  
& \frac{1}{2}(m_3 + m_1 - m_2) } .  
\end{eqnarray}  
Since the solar mixing angle is given in terms of $m_1$ and $m_2$ by Eq. (6),  
the maximal mixing of solar neutrinos implies $m_1=m_2$.  
However, it is not easy to naturally generate $\Delta m^2_{sol}$.  
For almost bimaximal mixing case \cite{bimax} which is due to nearly maximal   
solar mixing, similar to $\epsilon\neq 0$ case,   
neutrino mass spectrum is predicted  
providing $\sin^2 2\theta_{sol}$ is fixed so that the tiny mass splitting  
between $m_1$ and $m_2$ is naturally come out.  
  
At this stage, we address whether the above types of neutrino mass matrices   
can be obtained from some natural models of lepton masses and mixings.  
We will, first of all, show that the forms of the mass matrix given by Eqs. 
(9,12) can naturally be generated  
from some class of GUT models through a seesaw mechanism \cite{bando}.  
As shown in Refs. \cite{bando,dirac}, the relevant 
form of the Dirac neutrino  mass matrix is given in a parallel way with 
the up-type quark mass matrix in the GUT framework, 
\begin{eqnarray}  
m_U \simeq m \left(\begin{array}{ccc}  
0 & 0  &  r \\  
0 & r & 0 \\  
r & 0 & 1 \end{array} \right), 
\end{eqnarray} 
where $r$ is a small parameter of order $\sim (1/100 - 1/300)\sim m_c/m_t$ 
and the scale $m \sim m_t/3$ where the factor $3$ represents the effect of 
renormalization group equation. 
Then, the forms of the mass matrix given by Eqs. (9,12) can be obtained
through  the seesaw relation, $M_{\nu} = m_{U}^{T} M_R^{-1} m_{U}$, by taking 
the following forms of the right-handed majorana mass matrix: 
\begin{eqnarray}  
M_R \simeq  \left(\begin{array}{ccc}  
 aM^{\prime} & bM^{\prime} & 0 \\  
 bM^{\prime} & cM^{\prime} & 0 \\  
 0 & 0 & M \end{array}\right),  
~~~ 
M_R \simeq  \left(\begin{array}{ccc}  
 M^{\prime} & 0 & aM \\  
 0 & 0 & bM \\  
 aM & bM & cM \end{array}\right), 
\end{eqnarray}  
where the former corresponds to Eq. (9) and the latter to Eq. (12). 
We note that the non-diagonal form of the charged lepton mass
matrix diagonalized by a mixing matrix with very small off-diagonal 
elements does not hurt the form of $M_{\nu}$.
In order for the GUT scenario to be consistent with  the observed
bottom-tau mass ratio, it is required that the right-handed Majorana
mass of the third generation must be heavier than at least
$10^{13}$ GeV \cite{bando,bt}. 
Using the previous numerical results for the three neutrino masses,  
one can estimate the ranges of the parameters $(a,b,c)$ and $(M, M^{\prime})$
and then check whether the scales $M$ and $M^{\prime}$ are compatible with
the GUT scenario. 
While the consistent case for the former happens at the vacuum angle solution 
whics gives $M\sim 10^{10}$ GeV, $M^{\prime}\sim 10^{14}$ GeV, the consistent
cases for the latter happen at the small mixing and large mixing
angle MSW solutions which provide 
$M\sim 10^{10-11}, M^{\prime}\sim 10^{15-16}$ GeV.  
Here, the magnitude of $a,b,c$ are determined to be of order ${\cal O}(1)$,
and the scale $M^{\prime}$ is given by the order of around GUT scale.
Since an intermediate scale $10^{10-11}$ GeV is also naturally viable
in the GUT scenario, we may say that the above solutions we found are
quite natural.
On the other hand, while the form   
given by Eq. (10) can not be obtained in such a natural way through the 
canonical seesaw mechanism, it can, as shown in Ref. \cite{typeii}, be   
achieved in the type II seesaw model with approximate   
$L_e - L_{\mu} - L_{\tau}$.

To justify above ansatz that leads to the proposed form of neutrino mass  
matrix  with three texture zeros and neutrino spectrum,  
the precise determination of $U_{e3}$ element as well as the precise  
experiment for neutrinoless double beta decay will may be essential,   
which requires several oscillation channels to be probed at the same time.  
{}From the fact that the $\nu_{\mu} \rightarrow  
\nu_{\tau}$ disappearance channel is sensitive only to $|U_{\mu3}|^2$ and  
the $\nu_{\mu} \rightarrow \nu_{e}$ appearance channel is sensitive to  
the product $|U_{\mu 3}|^2 \cdot |U_{e3}|^2$,  one can determine $|U_{e3}|$  
by combining the regions to be probed in both channels.  
K2K \cite{k2k} will be expected to perform this, but it does not, at present,  
sensitivity in the $\nu_{\mu} \rightarrow \nu_{e}$ appearance channel to  
probe the region of $|U_{e3}|^2$ allowed by Super-Kamiokande, CHOOZ,  
and Palo Verde \cite{lisi}.  
  
In summary, we have examined some patterns of Majorana neutrino mass matrix   
which is compatible with the phenomenological lepton flavor mixing matrix and   
non-observation of neutrinoless double beta decay. We constructed  the lepton   
mixing matrix by taking $\theta_{23}=\pi/4$ which corresponds to the maximal   
mixing of the atmospheric neutrinos, $\theta_{12}=\theta_{sol}$, and allowing   
non-vanishing very small mixing angle $\theta_{13}$.   
Imposing $(M_{\nu})_{ee}=0$  
for the Majorana neutrino mass matrix {\it in the leading order},  
a relationship between the solar mixing angle and the ratio of the first  
two neutrino mass eigenvalues $m_1/m_2$ has been obtained.   
Additional possible texture zeros have been assigned to the mass matrix so as  
for the nonvanishing $\theta_{13}$ to be predictable in terms of neutrino   
masses.  
We have showed how three neutrino mass eigenvalues can be estimated  
from the relation for the solar mixing angle and the experimental results  
of $\Delta m^2_{sol}$ and $\Delta m^2_{atm}$ in this framework.  
We have also discussed how some forms of the mass matrix found in this paper  
can be achieved in any natural model of lepton masses.  
  
\acknowledgements  
  
\noindent  
We thank V. Barger and K. Whisnant for careful reading of the manuscript and  
their valuable comments.  
The work of C.S.K. was supported  in part by BK21 Project,  
in part by  Grant No. 2000-1-11100-003-1 and SRC Program of the KOSEF,  
and in part by the KRF Grants (Project No. 1997-011-D00015 and   
£ Project No. 2000-015-DP0077).  
\\  
  
{\it Note added:} while this paper is being completed,  we heard news of  
recent analysis for solar neutrino oscillation by SuperKamiokande which  
indicates that LMA is favored at $95\%$CL, whereas  
SMA and VO are disfavored. If it will be   
conformed in the future, only the part of LMA  
in our work is relevant.  
  

\end{document}